\documentclass{ws-ijmpa}
\usepackage[super,compress]{cite}
\usepackage{graphicx}
\usepackage{graphicx}
\usepackage{dcolumn}
\usepackage{bm}
\usepackage{epstopdf}
\newcommand{\hs}{\hspace*{0.5cm}}

\newcommand{\be}{\begin{equation}}
\newcommand{\ee}{\end{equation}}
\newcommand{\bea}{\begin{eqnarray}}
\newcommand{\eea}{\end{eqnarray}}
\newcommand{\ben}{\begin{enumerate}}
\newcommand{\een}{\end{enumerate}}
\newcommand{\bde}{\begin{widetext}}
\newcommand{\ede}{\end{widetext}}
\newcommand{\nn}{\nonumber}
\newcommand{\crn}{\nonumber \\}

\newcommand{\al}{\alpha}
\newcommand{\la}{\lambda}

\newcommand{\ga}{\gamma}

\newcommand{\om}{\omega}

\newcommand{\fr}{\frac}
\newcommand{\bc}{\begin{center}}
\newcommand{\ec}{\end{center}}
\newcommand{\Ga}{\Gamma}
\newcommand{\de}{\delta}

\newcommand{\ep}{\epsilon}

\begin{document}
%
\catchline{}{}{}{}{}

\title{NEUTRINO MIXING WITH NON-ZERO $\theta_{13}$ IN ZEE-BABU MODEL}

\author{HOANG  NGOC  LONG}

\address{Institute of Physics,
VAST, 10 Dao Tan, Ba Dinh, Hanoi, Vietnam\\
hnlong@iop.vast.ac.vn}

\author{VO VAN VIEN }

\address{Department of Physics, Tay Nguyen University, 567 Le
Duan, Buon Ma Thuot, Vietnam\\
wvienk16@gmail.com}
\maketitle

\begin{history}
\received{Day Month Year}
\revised{Day Month Year}
\end{history}

\begin{abstract}
The exact solution for the neutrino mass matrix of the Zee-Babu model
is derived. Tribimaximal mixing imposes conditions on the
Yukawa couplings, from  which the normal mass hierarchy is
preferred. The derived conditions give a possibility of Majorana maximal $\mathrm{CP}$ violation in the neutrino sector.
 We have shown that non-zero $\theta_{13}$ is generated if Yukawa couplings between leptons almost equal to
  each other. The model gives some regions of the parameters where neutrino mixing angles and the normal
   neutrino mass hierarchy obtained consistent with the recent experimental data.

\keywords{Neutrino mass and mixing, Non-standard-model neutrinos,
Zee-Babu model}
\end{abstract}

\ccode{PACS numbers: 14.60.Pq, 14.60.St}


\section{\label{intro}Introduction}

Nowadays, particle physicists are attracted by two exciting
subjects: Higgs and neutrino physics. The neutrino mass and mixing
are the first evidence of beyond Standard Model physics. Many
experiments show that neutrinos have tiny masses and their mixing
is sill mysterious \cite{altar1, altar2}.  Recent data are a clear
sign of rather large value $\theta_{13}$ \cite{smirnov} .

The tribimaximal  (TBM) form for explaining the lepton
mixing scheme was first proposed by Harrison-Perkins-Scott (HPS),
which apart from the phase redefinitions, is given by
\cite{hps1,hps2}
\begin{eqnarray}
U_{\mathrm{HPS}}=\left(
\begin{array}{ccc}
\,\,\,\,\frac{2}{\sqrt{6}}       &\,\,\,\,\frac{1}{\sqrt{3}}  &0\\
-\frac{1}{\sqrt{6}}      &\,\,\,\,\frac{1}{\sqrt{3}}  &\frac{1}{\sqrt{2}}\\
\,\,\,\,\frac{1}{\sqrt{6}}      &-\frac{1}{\sqrt{3}}  &\frac{1}{\sqrt{2}}
\end{array}\right),\label{Utbm}
\end{eqnarray}
can be considered  as a good approximation for the recent
neutrino experimental data, where the large mixing angles are completely different from the
quark mixing ones defined by the Cabibbo-Kobayashi-Maskawa (CKM)
matrix \cite{pmns1, pmns2, pmns3}   .

The most recent
fits suggest that one of the mixing angles is approximately zero
and another has a value that implies a mass eigenstate
that is nearly an equal mixture of $\nu_\mu$ and $\nu_\tau$.
The parameters of neutrino oscillations such as the squared mass
differences and mixing angles are now very constrained. The data
in PDG2010 \cite{PDG2010} imply
\bea &&\sin^2(2\theta_{12})=0.87\pm 0.03,\hs
\sin^2(2\theta_{23})> 0.92,\hs
 \sin^2(2\theta_{13})<0.15,\crn
&& \Delta m^2_{21}=(7.59\pm0.20)\times 10^{-5} \mathrm{eV}^2,\hs
\Delta m^2_{32}=(2.43\pm0.13)\times
10^{-3}\mathrm{eV}^2,\label{PDG10}\eea where (and hereafter) the
best fits are taken into accounts. Whereas, the new data
\cite{PDG2012, valle, abe, adam,fogli, An, Ahn, valle1,
fogli1,lthem} have been given to be slightly modified from the old
fits (\ref{PDG10}): \bea &&\sin^2(2\theta_{12})=0.857\pm 0.024 ,
\hs\sin^2(2\theta_{13})=0.098\pm 0.013, \hs \sin^2(2\theta_{23})>
0.95, \crn && \Delta m^2_{21}=(7.50\pm0.20)\times 10^{-5}
\mathrm{eV}^2,\hs \Delta m^2_{32}=(2.32^{+0.12}_{-0.08})\times
10^{-3}\mathrm{eV}^2.\label{PDG12}\eea

On the other hand, the discovery of the the long- awaited Higgs boson at around 125 GeV
\cite{cms} opened a new chapter in particle physics. It is
essential for us to determine which model the discovered  Higgs
boson belongs to?  For this aim, the diphoton decay of the Higgs
boson plays a very important role. It is expected that new physics
might enter here to modify the standard model (SM) Higgs property.

For the above mentioned reasons, the search for an extended model
coinciding with the current data on neutrino and Higgs physics is
one of top our priorities. In our opinion, the model with the
simplest particle content is preferred. In the SM, neutrinos are
strictly massless. For neutrino mass, an original model pointed
out by Zee in Ref.\cite{zee}
 in which new scalars are added in the Higgs sector
with neutrino masses induced at the one-loop level. After that a
two-loop scenario called the Zee-Babu model \cite{zeebabu} was
proposed. The Zee-Babu model \cite{zee,zeebabu,babu} with just two
additional charged Higgs bosons ($h^-, k^{--}$) carrying lepton
number 2, is very attractive\footnote{In the recent paper
\cite{Daniel}, the parameter space of the model under
consideration has been reanalyzed, and the lower bounds for masses
of the singly and doubly charged Higgses lie between  1 to 2
TeV.}. In this model, neutrinos get mass from two-loop radiative
corrections, which can fit current neutrino data. Moreover, the
singly and doubly charged scalars that are new in the model can
explain the large annihilation cross section of a dark matter pair
into two photons  as hinted-at by the recent analysis of the Fermi
$\ga$-ray
  space telescope data \cite{bks} , if the new charged scalars are relatively light
 and have large couplings to a pair of dark matter particles. These new
  scalars can also enhance the $ B (H \rightarrow \ga \ga)$, as the
 recent LHC results may suggest.

The Zee-Babu model contains the Yukawa couplings which are
specific for lepton number violating processes. There has
 been much work \cite{Sierra1, ageng, nebot, Daniel, Garcia} constraining
the parameter space of the model, however
 the explicit values of
 neutrino masses and mixings have not been considered.

In this paper, starting from the neutrino mass matrix, we get the
exact solution, i.e., the eigenstates and the eigenvalues. As a
consequence, the neutrino mixing matrix follows. With this
exact solution, we can fit current data and get constraints
on the couplings. We hope that experiments in the near
future will  approve or rule out the model.

\section{Neutrino mass matrix in the Zee-Babu model}

The Zee-Babu model \cite{zeebabu} includes two $\mbox{SU}(2)_L$
singlet Higgs fields, a singly charged field $h^-$ and a doubly
charged field $k^{--}$. Moreover, right-handed neutrinos are not
introduced. The addition of these singlets gives rise to the
Yukawa couplings: \be {\cal L}_Y =
f_{ab}\overline{(\psi_{aL})^C}\psi_{bL}h^+ +
h^\prime_{ab}\overline{(l_{aR})^C}l_{bR}k^{++}+H.c.,\label{pt151}
\ee where $\psi_L$ stands for the left-handed lepton doublet,
$l_R$ for the right-handed charged lepton singlet and ($a, b = e,
\mu, \tau$) being the generation indices, a superscript $^C$
indicating charge conjugation. Here $\psi^C=C\overline{\psi}^T$
with $C$ being the charge-conjugation matrix. The coupling
constant $f_{ab}$ is antisymmetric ($f_{ab}=-f_{ba}$), whereas
$h_{ab}$ is symmetric ($h_{ab}=h_{ba}$). Gauge invariance
precludes the singlet Higgs fields from coupling to the quarks. In
terms of the component fields, the interaction Lagrangian is given
by\bea
 {\cal L}_Y&=&2\left[ f_{e\mu}(\bar{\nu_e^c}\mu_L - \bar{\nu_\mu^c}e_L) +
  f_{e\tau}(\bar{\nu_e^c}\tau_L - \bar{\nu_\tau^c}e_L) +
   f_{\mu \tau}(\bar{\nu_\mu^c}\tau_L - \bar{\nu_\tau^c}\mu_L)\right]h^+\crn
&+&\left[h_{ee}\bar{e^c}e_R + h_{\mu \mu}\bar{\mu^c}\mu_R +h_{\tau \tau}
\bar{\tau^c}\tau_R + h_{e \mu}\bar{e^c}\mu_R + h_{e \tau}\bar{e^c}\tau_R
+ h_{\mu \tau}\bar{\mu^c}\tau_R\right]k^{++} \label{pt151t}\\
&+&H.c. \nn\eea
where we have used $h_{aa} = h^\prime_{aa}, h_{ab} = 2 h^\prime_{ab} $ for
$a \neq b$.
Eq. (\ref{pt151}) conserves
lepton number,  therefore, itself cannot be responsible for
neutrino mass generation.

The Higgs potential contains the terms:
\be V(\phi,h^+,k^{++})=\mu(h^- h^- k^{++}+h^+ h^+
k^{--})+\cdot\cdot\cdot, \label{pt152}\ee which violate lepton
number by two units. They are expected to cause Majorana neutrino
masses.

\begin{figure}
\bc
\includegraphics{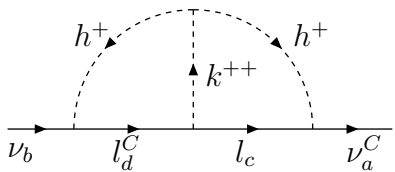}
\caption{\label{Surdiad} The two-loop diagram in the Zee-Babu
model.} \ec
\end{figure}

In the literature, Majorana neutrino masses are
generated at the two-loop level via the diagram shown in
\cite{babu} and again depicted in Fig.\ref{Surdiad}.
The corresponding mass matrix  for Majorana neutrinos is  as
follows \bea M_{ab}&=&8\mu
f_{ac}h^*_{cd}m_cm_dI_{cd}(f^+)_{db}.\label{pt152t}\eea The
integral $I_{cd}$ is given by \cite{tinhI}
 \bea
I_{cd}&=&\int\fr{d^4k}{(2\pi)^4}\int\fr{d^4q}{(2\pi)^4}\fr{1}{k^2-m^2_c}
\fr{1}{k^2-M^2_h}\fr{1}{q^2-m^2_d}\crn
&\times&\fr{1}{q^2-M^2_h}\fr{1}{(k-q)^2-M^2_k} \label{pt153}\eea
Note that Eq. (\ref{pt153}) can be simplified by neglecting the
charged lepton masses $m_c$ and $m_d$ \cite{babum} .

To evaluate the integral  above, one neglects the charged lepton
masses in the denominator, since these masses are much smaller
than the charged
  scalar masses $M_h$ and $M_k$. Then
\bea I_{cd} \simeq I = \frac{1}{(16\pi^2)^2}\frac{1}{M^2}
\fr{\pi^2}{3} \tilde{I}(r), \,  \, M \equiv  \textrm{max}(M_k,M_h)
\label{pt154}\eea which does not depend on lepton masses. Here
$\tilde{I}(r)$ is a function of the ratio of the masses of the
charged Higgses $r \equiv M^2_k/M^2_h$,
\be \tilde{I}(r) =\left\{%
\begin{array}{ll}
     \hbox{$1 + \fr{3}{\pi^2}(\log^2 r - 1)$} & \mbox{for} \, \, r \gg 1 \\
\hbox{$1$} & \mbox{for} \, \,  r \rightarrow 0, \\
\end{array}%
\right.\label{lv1}\ee which is close to $1$ for a wide range of
scalar masses.

The neutrino mass matrix arising from (\ref{pt152t}) is symmetric and given by
\bea &&{\cal M}_{\nu} = - I \mu f_{\mu \tau}^2 \times\crn
 &&\left(%
\begin{array}{ccc}
  \ep^2 \om_{\tau \tau } + 2 \ep \ep^\prime \om_{\mu \tau} + \ep^{\prime 2} \om_{\mu \mu}\,\,\,
   & \ep \om_{\tau \tau} +  \ep^\prime( \om_{\mu \tau} - \ep  \om_{e \tau}
 -\ep^{\prime } \om_{e \mu}) &\,\,\,-\ep^\prime \om_{\mu \mu}  -  \ep( \om_{\mu \tau}+  \ep \om_{e\tau}
 +   \ep^\prime \om_{e\mu}) \\
  \star &  \om_{\tau \tau} + \ep^{\prime 2} \om_{e e} -2\ep^\prime \om_{e\tau} &
   \ep \ep^\prime \om_{ee} - \om_{\mu\tau}-\ep \om_{e\tau}  +\ep^\prime \om_{e\mu} \\
  \star & \star &  \om_{\mu \mu} +2 \ep \om_{e\mu} + \ep^2 \om_{ee} \\
\end{array}%
\right)\crn \label{ktra}\eea
where we have redefined parameters:
\bea \ep \equiv \fr{f_{e\tau}}{f_{\mu \tau}}, \hs  \ep^\prime
\equiv \fr{f_{e\mu}}{f_{\mu \tau}} \hs \om_{a b} \equiv m_a
h^*_{ab} m_b .\label{ef}\eea

Let us denote \bea \om_{\tau\tau}^{'} &\equiv&
 \om_{\tau\tau}
+ \ep^{\prime 2} \om_{e e} -2\ep^\prime \om_{e\tau},
 \crn
\om_{\mu\tau}^{'} &\equiv&   \om_{\mu\tau}+\ep
\om_{e\tau}  -\ep^\prime \om_{e\mu} -\ep \ep^\prime \om_{ee} , \\
\om_{\mu\mu}^{'} &\equiv& \om_{\mu \mu} +2 \ep \om_{e\mu} + \ep^2
\om_{ee}\nn, \eea then the neutrino mass matrix can be rewritten
in the compact form
 \bea
 {\cal M}_{\nu} = - I \mu f_{\mu \tau}^2\left(%
\begin{array}{ccc}
  \ep^2 \om_{\tau \tau }^\prime + 2 \ep \ep^\prime \om_{\mu \tau}^\prime + \ep^{\prime 2} \om_{\mu \mu}^\prime
   & \ep  \om^\prime_{\tau \tau} +  \ep^\prime \om^\prime_{\mu \tau}  & -\ep  \om_{\mu\tau}^{\prime}
   -\ep^\prime \om_{\mu\mu}^{\prime} \\
  \star & \om_{\tau\tau}^{\prime} &
-   \om_{\mu\tau}^{\prime}  \\
  \star & \star &  \om^\prime_{\mu \mu} \\
\end{array}%
\right).\label{ktra}\eea
The above matrix has three exact eigenvalues given by \bea
m_1&=&0,\crn
 m_{2,3}&=&\frac{1}{2}\left(-k F \pm\sqrt{k^2\left[F^2+4(1+\epsilon^2+\epsilon'^2)(\om'^2_{\mu\tau}-\om'_{\mu\mu}
 \om'_{\tau\tau})\right]}\right),\label{m123}\eea
 where we have denoted
 \be
k=\mu If^2_{\mu\tau},\hs F=(1 + \epsilon'^2)\om'_{\mu\mu}+2\epsilon\epsilon'\om'_{\mu\tau}
+ (1 +\epsilon^2)\om'_{\tau\tau}.\label{kF} \ee
The massless  eigenstate  is given by \bea \nu_1 &=&
\fr{1}{\sqrt{f^2_{e\mu}+f^2_{e\tau}+f^2_{\mu\tau}}}
(f_{\mu\tau}\nu_e-f_{e\tau}\nu_\mu+f_{e\mu}\nu_\tau).\nn\\
\label{eigz}\eea
The mass matrix (\ref{ktra}) is diagonalized as \[ U^T {\cal
M}_{\nu} U=\mathrm{diag}(0, m_2, m_3), \] where
\bea
U&=&\left(\begin{array}{ccc}
 \,\,\, \frac{1}{\sqrt{1+\epsilon^2+\epsilon'^2}} & -\frac{A_1}{\sqrt{1+A^2_1+B^2_1}} & \frac{A_2}{\sqrt{1+A^2_2+B^2_2}} \\
 -\frac{\epsilon}{\sqrt{1+\epsilon^2+\epsilon'^2}} &  - \frac{B_1}{\sqrt{1+A^2_1+B^2_1}} &\frac{B_2}{\sqrt{1+A^2_2+B^2_2}} \\
 \,\,\,\frac{\epsilon'}{\sqrt{1+\epsilon^2+\epsilon'^2}} &- \frac{1}{\sqrt{1+A^2_1+B^2_1}} & \frac{1}{\sqrt{1+A^2_2+B^2_2}}
\end{array}\right)\label{massmix}
\eea
with the new notations
\bea
A_{1,2}&=&\frac{-k\left[\epsilon(\epsilon'^2-1)\om'_{\mu\mu}+2\epsilon'(1+\epsilon^2)
\om'_{\mu\tau}+\epsilon(1+\epsilon^2)\om'_{\tau\tau}\right]
\pm\ep\sqrt{k^2F'}}
{2k\left[\epsilon\epsilon'\om'_{\mu\mu}+(1+\epsilon^2)\om'_{\mu\tau}\right]},\label{gta}\\
B_{1,2}&\equiv
&\frac{k(1+\epsilon'^2)\om'_{\mu\mu}-k(1+\epsilon^2)\om'_{\tau\tau}\pm\sqrt{k^2F'}}
{2k\left[\epsilon\epsilon'\om'_{\mu\mu}+(1+\epsilon^2)\om'_{\mu\tau}\right]},
\label{gtb} \eea and
 \bea
F'=F^2+4(1+\epsilon^2+\epsilon'^2)(\om'^2_{\mu\tau}-\om'_{\mu\mu}
 \om'_{\tau\tau}).\label{Fp}
\eea
 The eigenstates  $\nu_i$ corresponding to the  eigenvalues
$m_i \, (i=1,2,3)$ are found to be \bea
\nu_1&=& \fr{1}{\sqrt{f^2_{e\mu}+f^2_{e\tau}+f^2_{\mu\tau}}}
(f_{\mu\tau}\nu_e-f_{e\tau}\nu_\mu+f_{e\mu}\nu_\tau),\crn
\nu_2&=&
-\frac{A_1}{\sqrt{1+A^2_1+B^2_1}}\nu_e-\frac{B_1}{\sqrt{1+A^2_1+B^2_1}}
\nu_\mu- \frac{1}{\sqrt{1+A^2_1+B^2_1}}\nu_\tau,\crn
\nu_3&=&
\frac{A_2}{\sqrt{1+ A^2_2+B^2_2}}\nu_e +\frac{B_2}{\sqrt{1+A^2_2+B^2_2}} \nu_\mu +
\frac{1}{\sqrt{1+A^2_2+B^2_2}}\nu_\tau.\label{root} \eea From the
explicit expressions of $A_{1,2}$ and $B_{1,2}$ in (\ref{gta}) and (\ref{gtb}), some useful relations are in
order
  \bea A_1A_2+B_1B_2+1&=&0,\crn A_1-\epsilon
B_1+\epsilon'&=&0,\crn A_2-\epsilon B_2+\epsilon'&=&0,\crn
(A_1-A_2)/(B_1-B_2)&=&\epsilon. \label{ABrelation}\eea
One also has
\bea
A_1A_2&=&\frac{(\epsilon'^2-\epsilon^2)\om'_{\mu\tau}
+\epsilon\epsilon'(\om'_{\tau\tau}-\om'_{\mu\mu})}{\epsilon\epsilon'\om'_{\mu\mu}+(1+\epsilon^2)\om'_{\mu\tau}},\crn
B_1B_2&=&-\frac{(1+\epsilon'^2)\om'_{\mu\tau}
+\epsilon\epsilon'\om'_{\tau\tau}}{\epsilon\epsilon'\om'_{\mu\mu}+(1+\epsilon^2)\om'_{\mu\tau}}.
\nn \eea

\section{\label{TBMconstraint}Constraints from the tribimaximal mixing}
The current data on neutrino mass and mixing show that
tribimaximal mixing \cite{hps1, hps2} as displayed in (\ref{Utbm})
is very specific. Comparing (\ref{massmix}) with (\ref{Utbm})
yields the following conditions \bea
\ep &=&\ep^\prime =\fr 1 2,\label{dhep}\\
A_2 &= &0, \hs A_1=B_1 =-1,\hs B_2 = 1\label{dkAB}
  \eea
Eqs. (\ref{dhep}) and (\ref{ef}) lead to \be
f_{e \mu} = f_{e \tau} =   \fr 1 2 f_{\mu \tau}\label{dhep2}
\ee
Substitution of (\ref{dhep}) into expressions of $A_{1,2}$,
$B_{1,2}$ in (\ref{gta}) and (\ref{gtb}) yields
\bea
A_{1,2}&=&\frac{k(3\om'_{\mu\mu}-10\om'_{\mu\tau}-5\om'_{\tau\tau})
\pm\sqrt{k^2F_0}}{4k(\om'_{\mu\mu}+5\om'_{\mu\tau})},\label{A1A2}\\
B_{1,2}&=&\frac{5k(\om'_{\mu\mu}-\om'_{\tau\tau})
\pm\sqrt{k^2F_0}}{2k(\om'_{\mu\mu}+5\om'_{\mu\tau})},\label{B1B2}
\eea with \bea F_0=4(\om'_{\mu\mu}+5\om'_{\mu\tau})^2
+(\om'_{\mu\mu}-\om'_{\mu\tau})(21\om'_{\mu\mu}-20\om'_{\mu\tau}-25\om'_{\tau\tau}).\label{F0}
\eea If $\om^\prime_{\mu\mu}=\om^\prime_{\tau\tau}=\om'$ we have:
\bea
A_{1,2}&=&-\frac{1}{2}\left(1\mp\frac{k(\om'+5\om'_{\mu\tau})}{\sqrt{k^2(\om'+5\om'_{\mu\tau})^2}}\right),\label{A12v}\\
B_{1,2}&=&\pm\frac{k(\om'+5\om'_{\mu\tau})}{\sqrt{k^2(\om'+5\om'_{\mu\tau})^2}}.\label{B12v}\eea
It can be checked that with the help of (\ref{dhep}), all
remaining conditions in (\ref{dkAB}) are satisfied if \be
\om^\prime_{\mu\mu}=\om^\prime_{\tau\tau}\equiv\om',\label{dkvien}\ee
and $k(\om'+5\om'_{\mu\tau})$ are negative real numbers. This can be equivalently converted into a relation among the
Yukawa couplings \be \om_{\mu\mu} +\om_{e\mu} = \om_{\tau\tau}-
\om_{e\tau} \label{dkv1}\ee Note that our derived constraints are
somewhat different from those given in \cite{ageng} .

From the conditions (\ref{dhep}) and (\ref{dkvien}) we obtain
\footnote{The integration in Fig.\ref{Surdiad} is linear divergent
and has a surface term \cite{donglongsurface} , which give a
similar form of mass matrix.}

\bea m_1&=&0, \crn
m_{2,3}&=&-\frac{1}{4}\left[k(5\om'+\om'_{\mu\tau})\mp\sqrt{k^2(\om'+5\om'_{\mu\tau})^2}\right].\label{m1230}
\eea
The complex phases which can arise when diagonalizing the
neutrino mass matrix (\ref{ktra}) can be
 absorbed by the redefinition of the mass matrix eigenvectors, as it should be given that both $m_{2,3}$ are physical
 observables. Hence, in this work we assume $m_2,m_3$ to be real.

Depending on the sign of the function in the square root
 we have two cases in which $k(5\om'+\om'_{\mu\tau})$ being either positive or negative. To fit the
  experimental data in \cite{PDG2010} the following condition must be satisfied
\bea
k(\om'+5\om'_{\mu\tau})<0. \label{Viencond1}\eea
The neutrino masses in (\ref{m123}) becomes \bea m_1&=&0,\hs m_2=-\frac{3k}{2}(\om'+\om'_{\mu\tau}),\hs
m_3=k(-\om'+\om'_{\mu\tau}).\label{m123v1} \eea
Taking the central values from the data \cite{PDG2010} as displayed in
(\ref{PDG10}), we have the two following solutions:
\begin{enumerate}
\item   $m_1=0, \hs m_2=0.008712 \,\, \mathrm{eV}, m_3=- 0.050059\,\,  \mathrm{eV}$ and then
\be U=
 \left(\begin{array}{ccc}
  \,\,\, \fr{2}{\sqrt{6}} & \,\,\,0.57735 & 4.17428\times 10^{-17}\\
  - \fr{1}{\sqrt{6}} &\,\,\,0.57735 & 0.707107 \\
   \,\,\,\fr{1}{\sqrt{6}} & -0.57735 & 0.707107
 \end{array}\right).\label{hpsv1}
\ee
In this case,  $\om'_{\mu\tau}$ and $\om'$ depend only on $k$ due to the following relations:
\bea
\om'_{\mu\tau}&=&-\frac{0.0279335}{k},\hs \om' =\frac{0.0221255}{k},\label{omv1}\\
\frac{\om'_{\mu\tau}}{\om'}&=&-1.2625 \label{omratiov1},\\
k(\om'&+&5\om'_{\mu\tau})=-0.117542<0.
\nn \eea
\item  $m_1=0, \hs m_2=-0.00871206\, \mathrm{eV}, m_3=-0.050059\,  \mathrm{eV}$, and
\be U=
 \left(\begin{array}{ccc}
   \fr{2}{\sqrt{6}} & 0.57735 &-5.93338\times 10^{-17}\\
  - \fr{1}{\sqrt{6}} &0.57735 & -0.707107 \\
   - \fr{1}{\sqrt{6}} & 0.57735 & 0.707107
 \end{array}\right).\label{hpsv2}
\ee
In this case $\om'_{\mu\tau}$ and $\om'$ depend only on $k$ according to the following relations:
\bea
\om'_{\mu\tau}&=&-\frac{0.0221255}{k},\hs \om' =\frac{0.0279335}{k},\label{omv2}\\
\frac{\om'_{\mu\tau}}{\om'}&=&-0.792076 \label{omratiov2},\\
k(\om'&-&5\om'_{\mu\tau})=-0.0826938<0.
\nn \eea
\end{enumerate}
The expressions (\ref{omratiov1}) and (\ref{omratiov2}) show that $\om'_{\mu\mu}, \om'_{\tau\tau}$ and $\om'_{\mu\tau}$
 are of the same order, and the normal neutrino mass hierarchy was used
 \footnote{Here, we have assumed a normal neutrino mass hierarchy in which $m_{1}=\lambda_1=0,
  m_{2}=\lambda_2, m_{3}=\lambda_3$ where $\lambda_{i} \, (i=1,2,3)$ are eigenvalues of
   $M_\nu$ in (\ref{ktra}). A spectrum with inverted ordering can be obtained by using
    the notation $m'_{3}=\la_1=0, m'_{2}=\la_3\equiv m_3$ and $m'_{1}=\la_2\equiv m_2$.}.

In terms of the usual neutrino-oscillation parameters, the
matrices (\ref{hpsv1}) and (\ref{hpsv2}) mean that \be \sin^2_{23}
=\frac{1}{2},\hs \sin^2_{12}=\frac{1}{3}, \hs \sin^2_{13} =
0.\label{tbm1} \ee which in good agreement with the
tribimaximal form\cite{PDG2010}.  However, with a vanishing
$\theta_{13}$ now excluded  at more than $10 \sigma$ \cite{valle1}
the situation has changed somewhat and the result in (\ref{PDG10})
should be considered just as a good approximation.

Using  the standard parametrization of the  neutrino  mixing matrix (the PMNS matrix)  in terms of three angles and
CP violating  phases \cite{pmns1, pmns2, pmns3}
\bea
U&=&
\left(\begin{array}{ccc}
 1 & 0 & 0 \\
 0 &  c_{23} & s_{23} \\
 0 & -s_{23} & c_{23}
\end{array}\right)
\left(\begin{array}{ccc}
 c_{13} & 0 & s_{13}\ e^{-i\de} \\
 0 &  1 & 0 \\
 -s_{13}\ e^{i\de} & 0 & c_{13}
\end{array}\right)
\left(\begin{array}{ccc}
  c_{12} & s_{12} & 0 \\
 -s_{12} &  c_{12} & 0 \\
 0 & 0 & 1
\end{array}\right)
\left(\begin{array}{ccc}
 1 & 0 & 0 \\
 0 &  e^{i \ga/2} & 0 \\
 0 & 0 & 1
\end{array}\right)\crn
&=&
\left(\begin{array}{ccc}
  c^{}_{12}
c^{}_{13} & s^{}_{12} c^{}_{13} & s^{}_{13} e^{-i\delta} \\
-s^{}_{12} c^{}_{23} - c^{}_{12} s^{}_{13} s^{}_{23} e^{i\delta} &
c^{}_{12} c^{}_{23} - s^{}_{12} s^{}_{13} s^{}_{23} e^{i\delta} &
c^{}_{13} s^{}_{23} \\
s^{}_{12} s^{}_{23} - c^{}_{12} s^{}_{13}
c^{}_{23} e^{i\delta} & -c^{}_{12} s^{}_{23} - s^{}_{12} s^{}_{13}
c^{}_{23} e^{i\delta} & c^{}_{13} c^{}_{23} \\
\end{array}\right) \left(\begin{array}{ccc}
 1 & 0 & 0 \\
 0 &  e^{i \ga/2} & 0 \\
 0 & 0 & 1
\end{array}\right).
 \label{Uparamet}
\eea
where  $\de$ and $\ga$ are the Dirac and Majorana CP phase, respectively,
and $s_{ij}=\sin\theta_{ij}, \, c_{ij}=\cos\theta_{ij}\,\, (ij =12,23,13)$.
The above Majorana
mass matrix is diagonalized by the PMNS matrix
\[ U^T  {\cal M}_\nu U=  {\cal M}_{diag} = {\rm  diag} (m_1, m_2, m_3).\]
In the case of the normal mass hierarchy, the four parameters
are described as \cite{ageng,babum}
\bea
&&\ep = \tan \theta_{12}
 \frac{s_{23}}{c_{13}}
+ \tan \theta_{13}\,  e^{i\de},\crn
&&\ep^\prime = \tan \theta_{12}
 \frac{s_{23}}{c_{13}}
- \tan \theta_{13}\,  e^{i\de},\crn
&&\fr{\om^\prime_{\mu\tau}}{\om^\prime_{\mu \mu}} =
-\frac{ c_{13}^2 s_{23}c_{23} }
     { c_{13}^2 c_{23}^2
     + r_{2/3} (s_{12}s_{13}c_{23}\ e^{-i\de} + c_{12}s_{23} )^2 e^{-i\ga} }
\label{pt158} \\
&&\hspace{1.2cm}
-\frac{ r_{2/3} (s_{12}s_{13}c_{23}\ e^{-i\de} + c_{12}s_{23})
            (s_{12}s_{13}s_{23}\ e^{-i\de} - c_{12}c_{23}) e^{-i\ga} }
      { c_{13}^2 c_{23}^2
      + r_{2/3} (s_{12}s_{13}c_{23}\ e^{-i\de} + c_{12}s_{23} )^2 e^{-i\ga} },\crn
&&\fr{\om^\prime_{\tau\tau}}{\om^\prime_{\mu\mu}} =
\frac{ c_{13}^2 s_{23}^2
     + r_{2/3} (s_{12}s_{13}s_{23}\ e^{-i\de} - c_{12}c_{23})^2 e^{-i\ga} }
     { c_{13}^2 c_{23}^2
     + r_{2/3} (s_{12}s_{13}c_{23}\ e^{-i\de} + c_{12}s_{23} )^2 e^{-i\ga}
     },\label{dkrhl}
\eea with $r_{2/3}=m_2 / m_3$.

We can easily see that with the help of (\ref{dkvien}), Eq. (\ref{dkrhl}) is automatically satisfied.
On the other hand, from (\ref{pt158}) ones can find the values of $\ga$ corresponding to
those of $m_2, m_3$ as shown in Table \ref{gamavalues}, in which the values of $\ga$ is
approximately equal  to $\frac{\pi}{2}$. So the condition
(\ref{dkvien}) leads to Majorana maximal $CP$ violation: $\sin \ga_{CP}
\simeq 1$, as mentioned  in Ref. \citen{marep}. \bc
\begin{table}
\caption{The values of $\ga$ corresponding to $m_2, m_3$}
\vspace*{0.25cm}
\bc
\begin{tabular}{|c|c|c|c|}
\hline$m_2 \,[\mathrm{eV}]$&$m_3\, [\mathrm{eV}]$& $e^{-i\ga}$&$\ga
[\mathrm{rad}]$ \\ \hline
 $0.008712$&$-0.0500591$& $1.00002$& $\pi/2$
\\\hline $0.00871$&$-0.0500591$& $1.00001$& $\pi/2$ \\\hline
\end{tabular}
\ec
\label{gamavalues}
\end{table}
\ec

The recent considerations have implied $\theta_{13}\neq 0$, but small as given in Ref.\citen{PDG2012}.
A deviations from the tribimaximal form would be achieved with a non-zero value of $A_2$ and a small
difference of $\ep$ and $\ep'$ as shown in Section \ref{theta13}.

\section{\label{theta13}Experimental constraints with non-zero $\theta_{13}$}

The realistic neutrino mixing will be slightly deviated
 from the tribimaximal form.
This will be achieved with a very small value of $A_2$ and $\ep' \simeq \ep\simeq \frac{1}{2}$.
 With the help of (\ref{ABrelation}), the matrix $U$ in (\ref{massmix}) becomes
\bea
U&=&\left(\begin{array}{ccc}
 \,\,\, \frac{1}{\sqrt{1+\epsilon^2+\epsilon'^2}} & \frac{\ep^2+\ep'(\ep'+A_2)}{\sqrt{(1+\epsilon^2+
 \epsilon'^2)[\ep^2+(1+\ep^2)A^2_2+2A_2\ep'+\ep'^2]}} & \frac{A_2\ep}{\sqrt{(1+A^2_2)\ep^2+(A_2+\ep')^2}} \\
 -\frac{\epsilon}{\sqrt{1+\epsilon^2+\epsilon'^2}} &  \frac{\ep(1-A_2\ep')}{\sqrt{(1+\epsilon^2
 +\epsilon'^2)[\ep^2+(1+\ep^2)A^2_2+2A_2\ep'+\ep'^2]}}& \frac{A_2+\ep'}{\sqrt{(1+A^2_2)\ep^2+(A_2+\ep')^2}} \\
 \,\,\,\frac{\epsilon'}{\sqrt{1+\epsilon^2+\epsilon'^2}} & - \frac{A_2(1+\ep^2)
 +\ep'}{\sqrt{(1+\epsilon^2+\epsilon'^2)[\ep^2+(1+\ep^2)A^2_2+2A_2\ep'+\ep'^2]}} &
  \frac{\ep}{\sqrt{(1+A^2_2)\ep^2+(A_2+\ep')^2}}
\end{array}\right)\label{U1}
\eea
Combining (\ref{U1}) and (\ref{Uparamet}) we obtain:
\bea
t_{12}&=&\frac{U_{12}}{U_{11}}=\frac{\ep^2+\ep'(A_2+\ep')}{\sqrt{\ep^2+
(1+\ep^2)A^2_2+2A_2\ep'+\ep'^2}},\label{t12}\\
t_{23}&=&\frac{U_{23}}{U_{33}}=\frac{A_2+\ep'}{\ep}.\label{t23}
\eea
with $t_{i j}=\tan\theta_{i j}\,\, (i j =12, 23, 13)$.

Since $\ep$ and $\ep'$ close to each other, it can be assumed that
\be
\ep'=\al \ep \label{eep}
\ee
where $\al$ is a constant close to 1.

From the expressions (\ref{t12}), (\ref{t23}) and (\ref{eep}) we obtain the following relations:
\bea
t_{23}&=&-\frac{\al \ep^3 (1+t^2_{12})+\sqrt{\Ga}}{\al^2\ep^3-\ep(1+\ep^2)t^2_{12}},\label{t12t23relat1}\\
A_2&=&\frac{\ep^3\al(1+\al^2)-\al \ep t^2_{12}+\sqrt{\Ga}}{t^2_{12}(1+\ep^2)-\al^2\ep^2},\label{A2v1}
\eea
or
\bea
t_{23}&=&\frac{-\al \ep^3 (1+t^2_{12})+ \sqrt{\Ga}}{\al^2\ep^3-\ep(1+\ep^2)t^2_{12}}.\label{t12t23relat2}\\
A_2&=&\frac{\ep^3\al(1+\al^2)-\al \ep t^2_{12}-\sqrt{\Ga}}{t^2_{12}(1+\ep^2)-\al^2\ep^2},\label{A2v2}
\eea
where
\be
\Ga =\ep^2 t^2_{12}[1+(1+\al^2)\ep^2][(1+\al^2)\ep^2-t^2_{12}].\label{Ga}
\ee
Substituting $A_2$ from (\ref{A2v1}) into (\ref{U1}) yields
\be
U=\left(\begin{array}{ccc}
U_{11}& U_{12}&U_{13} \\
U_{21}& U_{22}&U_{23} \\
U_{31}& U_{32}&U_{33} \\
\end{array}\right),\label{U2}
\ee
with
\bea
U_{11}&=&\frac{1}{\sqrt{1+(1+\al^2)\epsilon^2}}\,, \hs U_{21}=-\frac{\ep}{\sqrt{1+(1+\al^2)\epsilon^2}}\crn
U_{31}&=&\frac{\al \ep}{\sqrt{1+(1+\al^2)\epsilon^2}}\,, \hs
U_{12}=-\frac{\ep\left[\ep t^2_{12}+(1+\al^2)\ep^3t^2_{12}+\al \sqrt{\Ga}\right]}{\sqrt{\ep^3[1+(1+\al^2)\ep^2]^2\Ga'}},\crn
U_{22}&=&\frac{\ep\left[\al^4\ep^4 -(1+\ep^2)t^2_{12}+\al^2\ep^2(1+\ep^2-t^2_{12})
+\al\ep\sqrt{\Ga}\right]}{\sqrt{\ep^3[1+(1+\al^2)\ep^2]^2\Ga'}},\crn
U_{32}&=&\frac{\al(1+\al^2)\ep^5+\al \ep^3 +(1+\ep^2)\sqrt{\Ga}}{\sqrt{\ep^3[1+(1+\al^2)\ep^2]^2\Ga'}}\,,\,\,
U_{13}=\frac{\ep\left[\al (1+\al^2)\ep^3 -\al\ep t^2_{12}+\sqrt{\Ga}\right]}{\sqrt{\ep^3[1+(1+\al^2)\ep^2]\Ga'}},\crn
U_{23}&=&-\frac{\al \ep^3 (1+t^2_{12})+\sqrt{\Ga}}{\sqrt{\ep^3[1+(1+\al^2)\ep^2]\Ga'}},\hs\hs\,\,\,\,\,\,\,
U_{33}=\frac{\ep[\al^2\ep^2-t^2_{12}(1+\ep^2)]}{\sqrt{\ep^3[1+(1+\al^2)\ep^2]\Ga'}}\,,\label{Uelements}
\eea
where
\be
\Ga'=(1-\al^2)\ep t^2_{12}+\ep^3 (1+\al^2)(\al^2+t^2_{12})+2\al\sqrt{\Ga}.\label{Gap}
\ee
We see that  the neutrino mixing matrix in (\ref{U2}) with the elements
 given in (\ref{Uelements})
depends only on three parameters $\al, \ep$ and $t_{12}$. It is easily to show that the model can  fit
the recent experimental constraints on the neutrino mixing angles. Indeed, by choosing $\al\in (0.98, 1.00),
\ep \in (0.50, 0.505)$ and taking the new data  given in \cite{PDG2012} with $t_{12}=0.691$, we obtain
\bea
&&U_{11}\in (0.814\div 0.818),\hs U_{12}\in -(0.563 \div0.566), \,\, U_{13}\in (0.010\div 0.140),\crn
&&U_{21}\in -(0.409\div 0.412),\,\,\, U_{22}\in -(0.380\div 0.460),\,\, U_{23}\in -(0.790\div 0.830),\crn
&&U_{31}\in (0.4025\div 0.410),\,\,\,\, U_{32}\in (0.680\div 0.720),\,\,\,\,\,\,\, U_{33}\in -(0.540 \div 0.600).\eea
or
\bea
U&=&\left(\begin{array}{ccc}
 0.814\div 0.818 & -(0.563 \div0.566) & 0.010\div 0.140 \\
 -(0.409\div 0.412)& -(0.380\div 0.460)&-(0.790\div 0.830) \\
 0.4025\div 0.410&0.680\div 0.720& -(0.540 \div 0.600)
\end{array}\right).\label{Uendv1}
\eea

It is interesting to note that the model-independent
parametrization of non-TBM structures based on deviations from the
reactor, solar and atmospheric  angles \cite{king} and on small
perturbations of the tribimaximal mixing eigenvectors
\cite{Sierra2} is similar to our approach here. Our  set
$\epsilon, \alpha, t_{12}$ is equivalent to the set
$\epsilon_{12}, \epsilon_{23}, \epsilon_{13}$ in Ref.
\cite{Sierra2}.

\begin{figure}[h]
\begin{center}
\includegraphics[width=13.0cm, height=4.0cm]{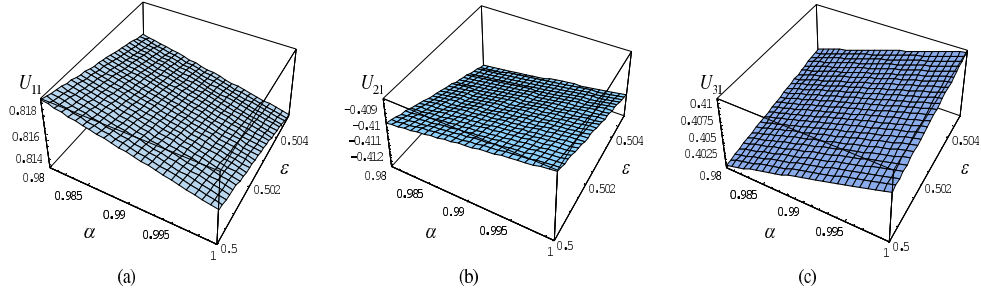}
\vspace*{-0.4cm} \caption[$U_{11}, U_{21}, U_{31}$ as functions of $\al$ and $\ep$ with $\al
\in (0.98, 1.00)$ and $ \ep \in (0.50, 0.505)$]{$U_{11}, U_{21}, U_{31}$ as functions of $\al$ and
$\ep$ with $\al \in (0.98, 1.00)$ and $ \ep \in (0.50, 0.505)$}\label{U1231v1}
\end{center}
\end{figure}

\begin{figure}[h]
\begin{center}
\includegraphics[width=13.0cm, height=4.0cm]{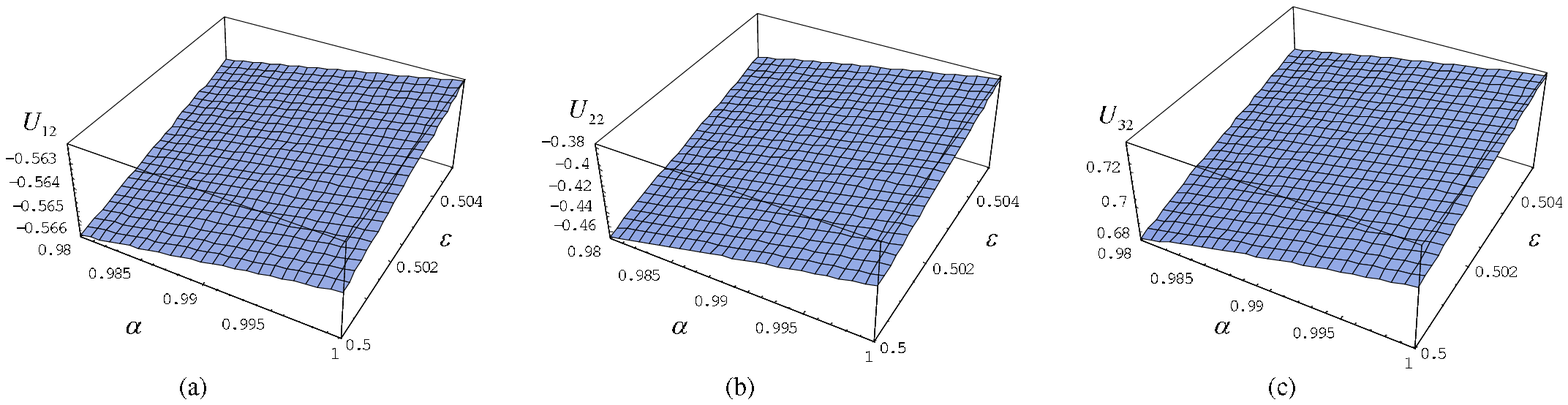}
\vspace*{-0.4cm} \caption[$U_{12}, U_{22}, U_{32}$ as functions of $\al$ and $\ep$ with $\al\in
(0.98, 1.00)$ and $ \ep \in (0.50, 0.505)$]{$U_{12}, U_{22}, U_{32}$ as functions of $\al$ and $\ep$
with $\al\in (0.98, 1.00)$ and $ \ep \in (0.50, 0.505)$}\label{U1232v1}
\end{center}
\end{figure}

\begin{figure}[h]
\begin{center}
\includegraphics[width=13.0cm, height=4.0cm]{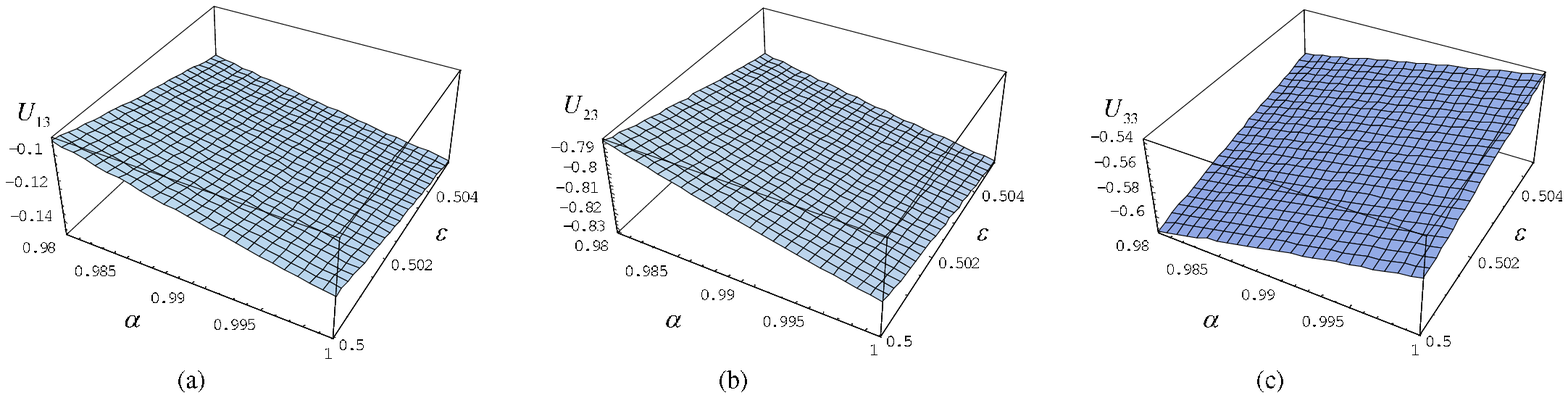}
\vspace*{-0.4cm} \caption[$U_{13}, U_{23}, U_{33}$ as functions of $\al$ and $\ep$ with $\al\in (0.98, 1.00)$
and $ \ep \in (0.50, 0.505)$]{$U_{13}, U_{23}, U_{33}$ as functions of $\al$ and $\ep$ with $\al
\in (0.98, 1.00)$ and $ \ep \in (0.50, 0.505)$}\label{U1233v1}
\end{center}
\end{figure}

 The Figs. \ref{U1231v1}a, \ref{U1231v1}b,
\ref{U1231v1}c, Figs.\ref{U1232v1}a, \ref{U1232v1}b,
\ref{U1232v1}c, Figs.\ref{U1233v1}a, \ref{U1233v1}b and
\ref{U1233v1}c give the dependence of the
 elements of $U$ matrix on $\al$ and $\ep$ with $t_{12}=0.691$.

\begin{figure}[h]
\begin{center}
\includegraphics[width=10.0cm, height=4.0cm]{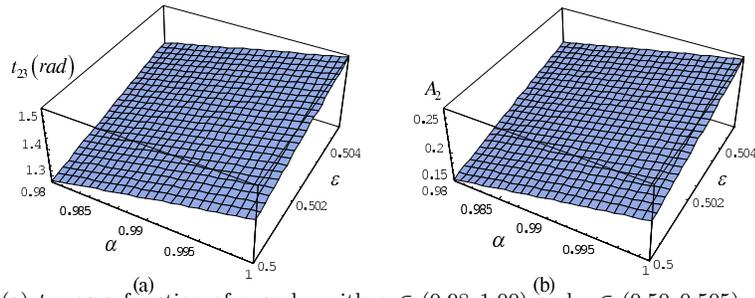}
\vspace*{-0.4cm} \caption[(a) $t_{23}$ as a function of $\al$ and $\ep$ with $\al\in (0.98, 1.00)$
and $ \ep \in (0.50, 0.505)$; (b) $A_{2}$ as a function of $\al$ and $\ep$ with $\al\in (0.98, 1.00)$
and $ \ep \in (0.50, 0.505)$]{(a) $t_{23}$ as a function of $\al$ and $\ep$ with $\al\in (0.98, 1.00)$
and $ \ep \in (0.50, 0.505)$; (b) $A_{2}$ as a function of $\al$ and $\ep$ with $\al\in (0.98, 1.00)$
and $ \ep \in (0.50, 0.505)$}\label{t23A2v1}
\end{center}
\end{figure}

With $\al\in (0.98, 1.00)$ and $ \ep \in (0.50, 0.505)$, from (\ref{t12t23relat1}) we obtain
 $t_{23}\in (1.3, 1.59)$ or $\theta_{23} \in (52.43^o, 56.31^o)$, and $A_2\in (0.15, 0.25)$ which
 are shown in Figs. \ref{t23A2v1}a and \ref{t23A2v1}b, respectively. In this case,
 $s_{13}\in (0.1, 0.14)$ or $\theta_{13}\in (5.74^o, 8.05^o)$.

Similarly, substituting $A_2$ from (\ref{A2v2}) into (\ref{U1}) yields
\bea
U&=&\left(\begin{array}{ccc}
 0.814\div 0.818 & 0.563 \div0.566 & -(0.010\div 0.140) \\
 -(0.409\div 0.412)& 0.69\div 0.73& 0.54\div 0.58 \\
 0.4025\div 0.410&-(0.38\div 0.44)& 0.8 \div 0.83
\end{array}\right)\label{Uendv2}
\eea provided that $\al\in (0.98, 1.00)$ and $\ep \in (0.50,
0.505)$. In this case, $t_{23}\in (0.65,0.75)$ or $\theta_{23}\in
(45^o, 50.19^o)$, $s_{13}\in (0.02, 0.08)$ or $\theta_{13}\in
(1.15^o, 4.6^o)$ and $A_2\in (0.05, 0.15)$. We note that in these
regions of the values of $\al$ and $\ep$, $\theta_{13}$ is smaller
than that given in \cite{PDG2012} , but the other regions of these
parameters will provide a consistent range of $\theta_{13}$, such
as, when $\al\in (0.98, 1.00)$ and $\ep \in (0.50, 0.51)$ then
$|s_{13}|\in (0.1, 0.16)$ or $\theta_{13} \in (5.74^o, 9.21^o)$.
This range of $\theta_{13}$ satisfies  the recent experimental
data in \cite{PDG2012} .

\begin{figure}[h]
\begin{center}
\includegraphics[width=13.0cm, height=4.0cm]{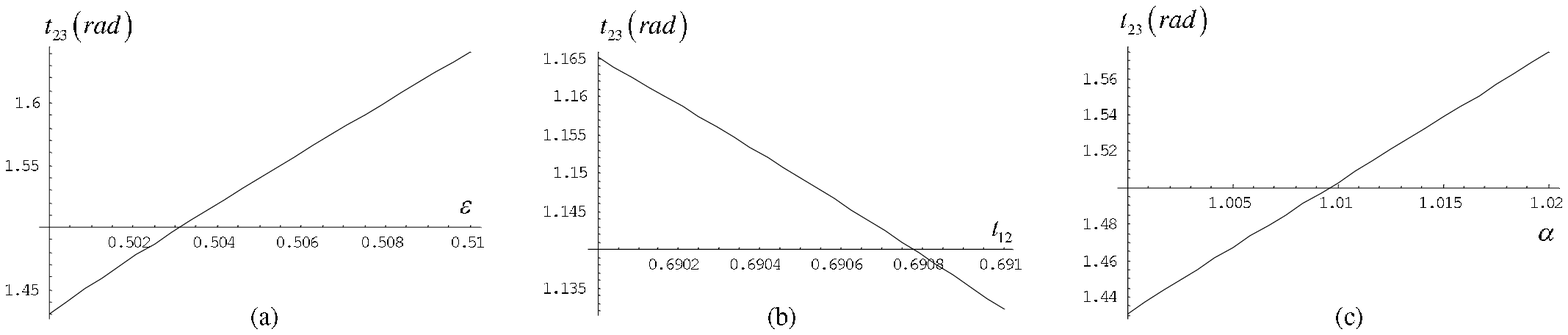}
\vspace*{-0.2cm}
\caption[(a) $t_{23}$ as a function of $\ep$ with $\ep\in(0.50, 0.51)$ and $\al=1, t_{12}=0.691$;
(b)  $t_{23}$  as a function of $t_{12}$ with $t_{12}\in (0.690, 0.691)$ and $\al=1,\, \ep=0.49$; (c)
  $t_{23}$ as a function of $\al$ with $\al \in (1.0, 1.02)$, $t_{12}=0.691,\, \ep=0.50$
   from (\ref{t12t23relat1})]{(a) $t_{23}$ as a function of $\ep$ with $\ep\in(0.50, 0.51)$
   and $\al=1, t_{12}=0.691$; (b)  $t_{23}$  as a function of $t_{12}$ with $t_{12}\in (0.690, 0.691)$
    and $\al=1,\, \ep=0.49$; (c)  $t_{23}$ as a function of $\al$ with $\al \in (1.0, 1.02)$, $t_{12}=0.691,\,
     \ep=0.50$ from (\ref{t12t23relat1}).}\label{t23t12ea}
\includegraphics[width=13.0cm, height=4.0cm]{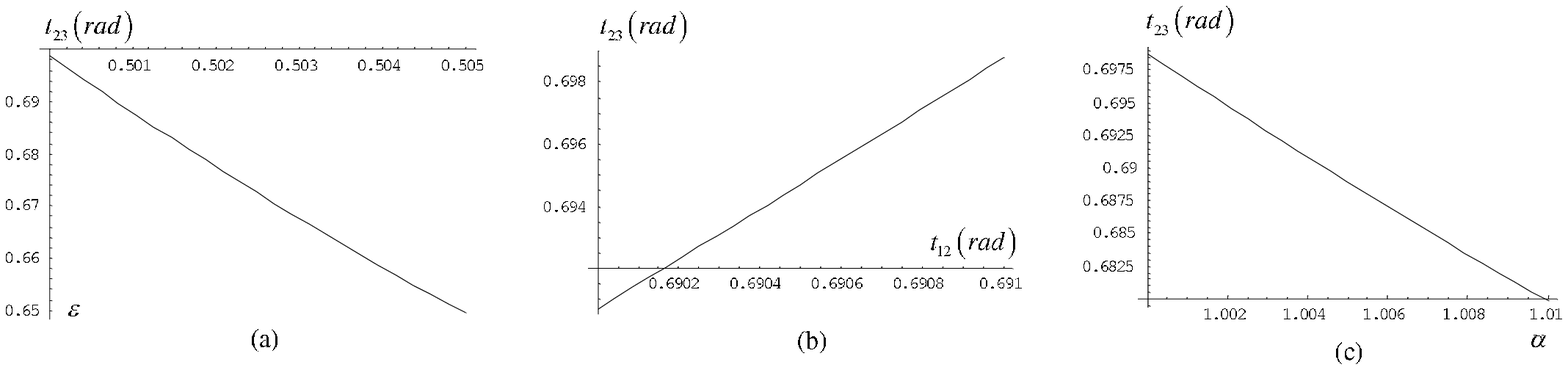}
\vspace*{-0.2cm}
\caption[(a) $t_{23}$ as a function of $\ep$ with $\ep\in(0.50, 0.505)$ and $\al=1, t_{12}=0.691$; (b)
$t_{23}$  as a function of $t_{12}$ with $t_{12}\in (0.690, 0.691)$ and $\al=1,\, \ep=0.50$; (c) $t_{23}$
as a function of $\al$ with $\al \in (1.00, 1.01)$, $t_{12}=0.691,\, \ep=0.50$ from
 (\ref{t12t23relat2})]{(a) $t_{23}$ as a function of $\ep$ with $\ep\in(0.50, 0.505)$ and
  $\al=1, t_{12}=0.691$; (b)  $t_{23}$  as a function of $t_{12}$ with $t_{12}\in (0.690,
  0.691)$ and $\al=1,\, \ep=0.50$; (c) $t_{23}$ as a function of $\al$ with $\al \in (1.00, 1.01)$,
   $t_{12}=0.691,\, \ep=0.50$ from (\ref{t12t23relat2}).}\label{t23t12eav1}
\end{center}
\end{figure}

From (\ref{t12t23relat1}) and  (\ref{t12t23relat2}) we can have the relations of $t_{23}$
and $t_{12}, \al, \ep$ as shown in the Figs. \ref{t23t12ea}a, \ref{t23t12ea}b and \ref{t23t12ea}c,
and  \ref{t23t12eav1}a, \ref{t23t12eav1}b and \ref{t23t12eav1}c, respectively, in which the values
of $\theta_{23}$ obtained encompass the best fit values in \cite{PDG2012} .

With the help of (\ref{eep}), $F$ in (\ref{kF}) becomes:
 \be
F=(1 + \al^2\ep^2)\om'_{\mu\mu}+2\al\ep^2\om'_{\mu\tau} + (1 +\epsilon^2)\om'_{\tau\tau},\label{kF1} \ee
and the physical neutrino masses from (\ref{m123}) is defined
 \bea
m_1&=&0,\crn
 m_{2,3}&=&\frac{1}{2}\left(-k F \pm\sqrt{k^2(F^2+B)}\right),\label{m123v1}\eea
 with
 \bea
 B=4[1+(1+\al^2)\ep^2](\om'^2_{\mu\tau}-\om'_{\mu\mu}\om'_{\tau\tau}).\label{B}
 \eea
Taking the central values from the data \cite{PDG2012} on neutrino mass square differences:
\bea
\Delta m^2_{21} = 7.50\times 10^{-5}\, \mathrm{eV^2},\hs \Delta m^2_{32} = 2.32\times 10^{-3}\, \mathrm{eV^2},\label{PDG12}
\eea
we obtain \bea
k&=&\frac{0.0402359}{F}.\label{kFrelation}
\eea
The neutrino masses are explicitly given as
\bea
&&m_1=0,\hs m_2=0.00871206\, \mathrm{eV},\hs m_3=-0.048948\, \mathrm{eV},\label{m123v1}\eea
which are in a normal ordering.

The ratio of $m_2$ to $m_3$ is given
\bea
&&\frac{|m_2|}{|m_3|}=0.177986\label{m23relat}
\eea
which is the same order as in Ref. \citen{ageng}.

\begin{figure}[h]
\begin{center}
\includegraphics[width=5.0cm, height=4.0cm]{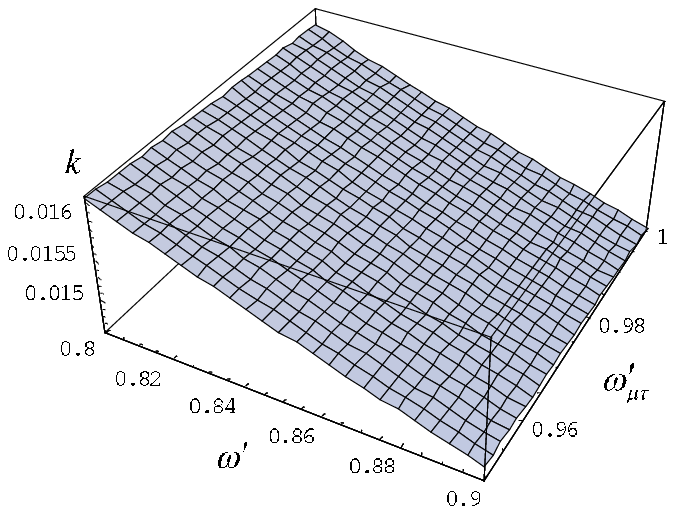}
\vspace*{-0.2cm} \caption[$k$ as a function of $\al, \ep, \om'$ and $\om'_{23}$ with $\om'_{\mu,\tau}\in
(0.95, 1.0)$, $\om'\in (0.80, 0.9)$ and $\al=1.00,\, \ep=0.5$]{$k$ as a function of $\al,
\ep, \om'$ and $\om'_{23}$ with $\om'_{\mu,\tau}\in (0.95, 1.0)$, $\om'\in (0.80, 0.9)$
 and $\al=1.00,\, \ep=0.5$}\label{kfig}
\end{center}
\end{figure}

Without loss of generality, we assume
$\om'_{\mu\mu}=\om'_{\tau\tau}=\om'$. From (\ref{kF}), (\ref{eep})
and (\ref{kFrelation}) we obtain the dependence of $k$ on $\al,
\ep$ and $\om$ and $\om'_{23}$: \bea k&=&\frac{0.0402359}{[(2 + (1
+ \al^2)\ep^2)]\om' + 2\al\ep^2\om'_{\mu\tau}}.\label{kaeom} \eea
 In the case $\al=1.00$ and $\ep=0.5$, one has
 \[k=-\frac{0.0402359}{2.5\om' + 0.5\om'_{\mu\tau}}.\]
 The Fig. \ref{kfig} gives the dependence of $k$ on $\om', \om'_{\mu\tau}$ with
 $\om'_{\mu,\tau}\in (0.95, 1.0)$ and $\om'\in (0.80, 0.9)$.

\section{Summary}

In this paper we have derived the exact eigenvalues and eigenstates of
the neutrino mass matrix in the Zee-Babu model. Tribimaximal
mixing
imposes some conditions on the Yukawa couplings. The constraints
derived in this work slightly differ from other ones given in the
literature, and the normal  mass hierarchy is preferred.
 The derived
conditions give a possibility of Majorana maximal $CP$ violation
in the neutrino sector.  We have shown that non-zero $\theta_{13}$
is generated if Yukawa couplings between leptons almost equal to
  each other. We have analyzed behaviors of the   mixing angles as
    functions of the Yukawa couplings, and
the model parameter space has been derived.

\section*{Acknowledgments}
This research is funded by the Vietnam National Foundation for Science
and Technology Development (NAFOSTED) under grant number
103.01-2011.63.

\end{document}